\documentclass[review, number]{elsarticle}
\graphicspath{ {./figures/} }
\usepackage[hidelinks]{hyperref}
\usepackage{float}
\usepackage{verbatim} 
\usepackage{amsmath}
\usepackage{multirow}
\usepackage{array}
\usepackage{booktabs}
\restylefloat{figure}
\floatstyle{plaintop} 
\restylefloat{table}
\usepackage[margin=0.80in]{geometry}

\usepackage{ccaption}
\usepackage{amsmath}
\usepackage{amsbsy}
\usepackage{dcolumn}
\usepackage{multirow}
\usepackage[table]{xcolor}
\usepackage{wrapfig}
\usepackage{pifont}
\usepackage{float}
\usepackage{adjustbox}
\usepackage{wrapfig}
\usepackage{soul}
\usepackage{footnote}
\usepackage{times}
\usepackage{lipsum}
\usepackage{lineno}

\usepackage[T1]{fontenc}
\usepackage{textcomp}
\usepackage{graphicx}
\usepackage{graphics}
\usepackage{soul}

\usepackage[font=normalsize,labelfont=bf]{caption}

\usepackage{epstopdf}

\usepackage{caption}
\usepackage{subcaption}

\usepackage{amssymb,amstext,amsmath}
\usepackage{float}
\usepackage{booktabs}
\usepackage{array,ragged2e}
\usepackage[table]{xcolor}
\usepackage{multirow}
\usepackage{algorithm}
\usepackage{algpseudocode}
\usepackage{pifont}
\usepackage{enumerate}

\usepackage{hyperref}

\begin{document}
\begin{frontmatter}











\title{Harnessing ADAS for Pedestrian Safety: A Data-Driven Exploration of Fatality Reduction}

\author[label1]{Methusela Sulle \corref{cor1}}
\ead{msulle@scsu.edu}

\author[label1]{Judith Mwakalonge}
\ead{jmwakalo@scsu.edu}

\author[label2]{Gurcan Comert}
\ead{gcomert@ncat.edu}

\author[label1]{Saidi Siuhi}
\ead{ssiuhi@scsu.edu}

\author[label1]{Nana Kankam Gyimah}
\ead{ngyimah@scsu.edu}

\cortext[cor1]{Corresponding author.}
\address[label1]{Department of Engineering, South Carolina State University, Orangeburg, South Carolina, USA, 29117}
\address[label2]{Department of Computational Engineering and Data Science, North Carolina A\&T State University, Greensboro, North Carolina, US, 27411}

\begin{abstract}
Pedestrian fatalities continue to rise in the United States, driven by factors such as human distraction, increased vehicle size, and complex traffic environments. Advanced Driver Assistance Systems (ADAS) offer a promising avenue for improving pedestrian safety by enhancing driver awareness and vehicle responsiveness. This study conducts a comprehensive data-driven analysis utilizing the Fatality Analysis Reporting System (FARS) to quantify the effectiveness of specific ADAS features like Pedestrian Automatic Emergency Braking (PAEB), Forward Collision Warning (FCW), and Lane Departure Warning (LDW), in lowering pedestrian fatalities. By linking vehicle specifications with crash data, we assess how ADAS performance varies under different environmental and behavioral conditions, such as lighting, weather, and driver/pedestrian distraction. Results indicate that while ADAS can reduce crash severity and prevent some fatalities, its effectiveness is diminished in low-light and adverse weather. The findings highlight the need for enhanced sensor technologies and improved driver education. This research informs policymakers, transportation planners, and automotive manufacturers on optimizing ADAS deployment to improve pedestrian safety and reduce traffic-related deaths.

\end{abstract}


\end{frontmatter}

\section{Introduction}
Traffic-related pedestrian injuries are a growing public health threat worldwide \cite{Chakravarthy2007}. More than a third of the 1.2 million people killed and the 10 million injured annually in road traffic crashes worldwide are pedestrians \cite{Crandall2002}. Improving pedestrian safety is key to encouraging walking because perceived traffic safety is a key determinant of walking for both transport and leisure \cite{Chong2022} and \cite{sulle2025unraveling}. A multifaceted approach is necessary, with design ideas emphasizing infrastructure improvements, driver education, and technological solutions \cite{Momin2025}. Despite notable advancements in vehicle safety technologies, pedestrian fatalities have continued to rise in recent years, presenting a paradox in traffic safety progress, the increasing number of pedestrian deaths can be attributed to multiple factors, increase in vehicle speed and traffic density, distracting pedestrians, reduces drivers' stopping rates \cite{Krizsik2024}, \cite{sulle2025vr} and \cite{Stimpson2013}. The widespread use of mobile phones while driving has become a serious public health threat and is linked to an increased risk of involvement in road crashes for both drivers and pedestrians, impairing situational awareness and reaction times \cite{Chengula2024}, \cite{MazharulHaque2023}.
Additionally, the growing popularity of larger and heavier vehicles, such as SUVs and pickup trucks, has exacerbated the problem \cite{JuliaKite-Laidlaw2024}. These factors underscore the importance of implementing targeted measures to enhance pedestrian safety \cite{Indie2024}.

One promising approach to enhancing pedestrian safety is the integration of Advanced Driver Assistance Systems (ADAS) in modern vehicles. ADAS is a set of such technologies that introduce an augmented layer of safety through cooperation between the driver and the vehicle \cite{Masello2022}. Key pedestrian-focused ADAS features include automatic emergency braking (AEB), pedestrian detection systems, lane departure warnings, adaptive cruise control, and forward-collision warnings \cite{NHTSA2022}. These systems rely on a combination of sensors, cameras, radar, and artificial intelligence to detect pedestrians and other road users in real time, enabling vehicles to automatically apply brakes or provide warnings to prevent a collision \cite{verizonconnect2024}. ADAS Technology can help reduce vehicle–pedestrian crashes, fatalities, and injuries \cite{Mahdinia2022}.  Aleksa reported that ADAS Automatic Emergency Braking has the greatest potential and could reduce 292 fatalities in the year 2030, while the Lane Departure Warning and the Curve-ABS for motorcycles would reduce 10 and 12 fatalities, respectively \cite{Aleksa2024}. However, ensuring safety measures for vulnerable road users (VRUs) such as pedestrians, cyclists, and e-scooter riders remains an area that requires more focused research effort \cite{AdnanYusuf2024}. This study has identified the following challenges:

\begin{enumerate}
    \item Human distractions (both pedestrian and driver) in fatal crashes: Drivers often engage with mobile phones and in-car entertainment systems, and pedestrians routinely use electronic devices, both of which reduce situational awareness \cite{Masello2023}.

    \item Effectiveness of ADAS under different conditions: Explore the effectiveness of ADAS under different conditions, such as nighttime, adverse weather, and rural areas with poor infrastructure \cite{Roh2020}. 
\end{enumerate}

To address the challenges outlined above, this study proposes a data-driven approach to enhance the effectiveness of Advanced Driver Assistance Systems (ADAS) in mitigating pedestrian fatalities and aims to contribute to:

\begin{enumerate}
    \item Assess the effectiveness of specific ADAS features in reducing pedestrian crashes.
        
    \item Explore environmental and behavioral factors influencing ADAS performance.

\end{enumerate}

The proposed method involves linking real-world crash data from national databases, i.e., the Fatality Analysis Reporting System (FARS), with vehicle-specific ADAS feature data to analyze the impact of these technologies under various environmental and behavioral conditions. The analysis also considers external conditions such as lighting and weather. The findings are aimed at informing policy recommendations, guiding ADAS technology development, and creating improved provisions for pedestrian protection in vehicle design and city planning.

The rest of this paper is organized as follows: Section $\text{II}$ presents related works on pedestrian safety. Section $\text{III}$ describes the proposed methodology. Section $\text{IV}$ describes the result and discussion. Finally, the conclusion and future work are provided in Section $\text{V}$.

\section{Related Works}
\subsection{Trends and Contributing Factors in Pedestrian Fatalities}
The probit model was employed to examine factors influencing the severity of intersection crashes. Key findings indicate that vehicle type, road type, collision type, driver characteristics, and time of day are all significant contributors to crash severity. Motorcycles, bicycles, and larger vehicles (trucks, buses) all elevate the threat of fatality, and nighttime and peak-hour crashes also are more likely to be severe. High speeds, intersections, and multi-vehicle collisions increase severity, whereas wet roads and turning maneuvers decrease it. Older drivers and young drivers, male drivers, and foreign-registered vehicles are all linked to higher crash severity. The study suggests the need for targeted safety measures, improved intersection design, and more driver awareness campaigns to reduce crash severity at intersections  \cite{TayRifaat2007}. Longitudinal Spatial Trends in U.S. Pedestrian Fatalities, 1999–2020, which analyzes the drastic 59\% increase in pedestrian fatalities since 2009 through spatial and socioeconomic trends. Based on Fatality Analysis Reporting System (FARS) data, U.S. Census data, and the Smart Location Database, the researchers mapped two decades of pedestrian fatalities. Their findings indicate a shift of fatalities from central city regions (which had a 63\% decline) to suburban regions (with a 32.1\% increase), particularly to post-war suburbs with low population and road densities. These high-risk areas often contain high minority populations, low education levels, and poverty levels over 60\% higher than the national average. The examination indicates the roles played by urban design and socioeconomic factors in pedestrian safety trends  \cite{Rodriguez2024}.
Furthermore, Crash trends in Texas (2010–2019) found that homelessness, income disparity, youth populations, and urbanization significantly influence crash rates. Crashes occur less often in rural locations but with a greater fatality risk due to speed and an absence of infrastructure. The primary factors include speed, nighttime, SUVs, and exurban development. The report calls for improved pedestrian infrastructure, lower speeds, and policy targeting to improve safety.
Understanding the trends and underlying causes of pedestrian fatalities is crucial in developing effective safety measures, such as the implementation of ADAS features in vehicles \cite{Bernhardt2021}.

\subsection{Advanced Driver Assistance Systems (ADAS)}
Advanced driver assistance systems (ADAS) are a set of such technologies that introduce an augmented layer of safety through cooperation between the driver and the vehicle \cite{Masello2022}. These systems use sensors, cameras, radar, and artificial intelligence to monitor the driving environment and provide real-time feedback or automated interventions \cite{Bathla2022}. The impact of ADAS was analyzed by using real accident data from the United States. Using a machine learning approach, i.e., Random Forest Classifier, the study identifies the contributing factors that influence injury severity in ADAS-equipped vehicles, i.e., mileage, month of the crash, and pre-crash speed. Results indicate that while ADAS features are effective in mitigating crashes, their performance worsens over time due to sensor degradation and also due to changing environmental conditions. Some of the issues raised by the study are data imbalance in the case of serious injury prediction and the need for frequent updates for ADAS technologies. Recommendations include enhancing system reliability, speed management functions, and data collection methods to optimize ADAS performance and road safety benefits in the long term  \cite{Okpono2024ADAS}.
Furthermore, the potential of ADAS in reducing road crashes and fatalities through the use of statistical crash data from Austria was examined. The study selects nine ADAS based on literature, expert interviews, and legislative trends, and it analyzes their potential for crash reduction by using a custom-designed software tool. Results show that braking and warning ADAS such as Forward Collision Warning (FCW) and Automatic Emergency Braking (AEB) hold the highest crash reduction potential, preventing an estimated 8,700 crashes and 70 fatalities by 2040. The report emphasizes proper use, driver education, and improved system reliability as key factors to achieve maximum ADAS benefits \cite{aleksa2024adas}.

\subsection{Effectiveness and Challenges of ADAS in Road Safety}
Using crash data from Victoria, South Australia, and Queensland (2013–2018), Peiris et al. \cite{Peiris2022} explored the impact of suboptimal road conditions on the effectiveness of Advanced Driver Assistance Systems (ADAS), particularly Autonomous Emergency Braking (AEB) and Lane Keep Assistance (LKA).The study estimates that 13–23\% of fatal and serious injury (FSI) crashes on arterial and sub-arterial roads are LKA-sensitive, but many benefits are lost due to poor road delineation. Annually, over 37 fatalities and 357 serious injuries across these states could remain unprevented due to inadequate infrastructure. The findings highlight the urgent need for road investments to fully support ADAS technologies and maximize their safety potential.

Besides, Advanced Driver Assistance Systems (ADAS) efforts towards enhancing pedestrian safety were assessed. Surveys and interviews with experts were employed to collect pedestrian and driver opinions on the effectiveness of ADAS technologies. The study confirms that ADAS is perceived generally as being beneficial although system feature preferences vary. Pedestrians and drivers show mixed acceptance of electronic detection methods, with concerns over privacy and feasibility. Experts highlight the need for greater sensor accuracy and defensive coding in autonomous vehicles\cite{ucha2021}.

The effectiveness of ADAS and Passive Safety Systems (PSS) in reducing road accidents from human errors was assessed by means of mathematical models in order to investigate interactions among drivers, ADAS warning and crash avoidance systems, and PSS. The study predicts the reliability of the vehicle under differing circumstances using reliability block diagrams. Results determine the possibility of increasing safety through ADAS but highlight challenges like system reliability and adaptation on the part of drivers. The research justifies vehicle design improvements and safety standards to enhance road safety \cite{Hojjati-Emami2012}. 
On the other hand, the feasibility of five ADAS functions—enhanced navigation, speed assistance, collision avoidance, intersection support, and lane keeping—as alternatives to traditional infrastructure-based safety measures. It finds that navigation and speed assistance are the most mature and cost-effective for large-scale implementation, while collision avoidance and intersection support still face technical and regulatory hurdles. The study highlights that radar, vision systems, and vehicle-to-vehicle communication show promise but require improvements in robustness, reliability, and cost. The integration of navigation with speed assistance could significantly enhance road safety, while other ADAS technologies need further development before widespread adoption \cite{Lu2005}.
Furthermore, crash-contributing factors among at-risk drivers using the SHRP 2 Naturalistic Driving Study and the potential of Advanced Driver Assistance Systems (ADAS) to mitigate these crashes. Findings show that driver error was responsible for 97\% of crashes, with recognition errors (56\%)—such as distraction and inadequate surveillance—being the most common. Teens and young adults exhibited more decision errors, while older adults had more performance errors. Automatic Emergency Braking (AEB) had the highest potential for crash mitigation (48\%), followed by vehicle-to-vehicle communication (38\%) and driver monitoring (24\%). The study suggests refining ADAS to take into account the weather conditions, improve the identification of non-compliant objects, and improve lane and pedestrian detection and giving critical guidelines to improve road safety through technology  \cite{Seacrist2021}.

While research on ADAS and pedestrian safety has been previously presented, there remain gaps in what is known about how ADAS function under different circumstances. This study fills these gaps through an exploratory examination of ADAS-equipped vehicle crashes using national crash data. Through analysis of environmental and behavioral conditions, the study sheds light on the way to optimize ADAS performance and inform pedestrian safety initiatives.

\section{Proposed Methodology}
The proposed methodology consists of two sections [1] data exploration and [2] experimental setting. Data exploration examines pedestrian crash trends, ADAS feature distribution, and contributing factors. The experimental setting includes VIN decoding to extract vehicle specifications to assess ADAS effectiveness under different conditions, such as lighting, driver distraction, and urban-rural environments.

\begin{figure*}
    \centering
    \includegraphics[width=0.9\linewidth]{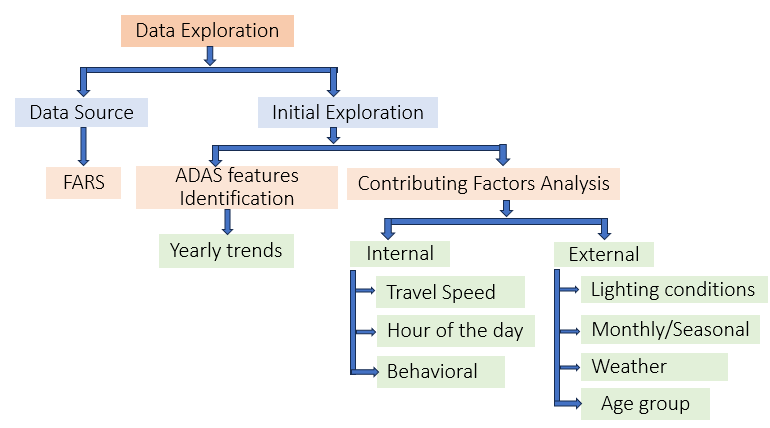}
    \caption{Exploratory Data Analysis}
    \label{fig:Data Exploration}
\end{figure*}

\begin{enumerate}
    \item \textbf{Data Exploration}
The study begins with data exploration to analyze pedestrian crash trends, ADAS feature distribution, and external factors influencing fatalities \ref{fig:Data Exploration}. This involves examining crash data from FARS to identify patterns in pedestrian-vehicle incidents, including variations by lighting conditions, urban versus rural environments, and driver or pedestrian distractions.

\item \textbf{Experimental Setting}: Vehicle Identification Number (VIN) decoding was done to identify vehicles with ADAS features such as Pedestrian Automatic Emergency Braking (PAEB), Forward Collision Warning (FCW), Lane Departure Warning (LDW), and Adaptive Cruise Control (ACC). This process follows NHTSA (National Highway Traffic Safety Administration) standards.
\end{enumerate}

\subsection{Data Analysis and Findings}
Descriptive statistical methods are used in the analysis to explore pedestrian accidents involving ADAS-equipped vehicles. Descriptive statistical methods enable us to gain insights into patterns, distributions, and trends of pedestrian fatalities. Some of the important measures are pedestrian accident trends annually, the proportion of cases involving ADAS-equipped and non-ADAS vehicles, and the distribution of accidents under different environmental conditions such as lighting and weather. Moreover, descriptive analysis examines behavior factors like driver distraction and pedestrian impairment to estimate their impact on ADAS performance.

\subsubsection{General Trends in Pedestrian Fatalities}
\begin{figure}
\centering
\centering
  \includegraphics[width=1\textwidth]{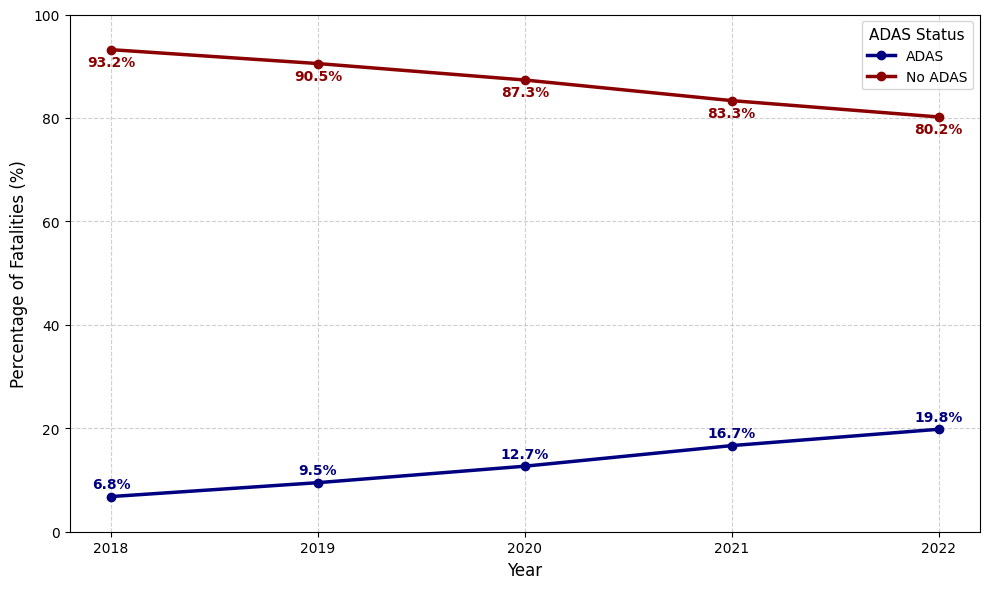}
\caption{Yearly Fatality trends by ADAS presence.}
  \label{fig:Yearly Trend}
\end{figure}

Figures \ref{fig:Yearly Trend} illustrate yearly trends and proportional involvement of Advanced Driver Assistance Systems (ADAS) in pedestrian fatalities from 2018 to 2022. The data reveals a consistent rise in pedestrian deaths involving ADAS-equipped vehicles, with a particularly sharp increase observed in 2021 and 2022. This finding is tied with Department of Transportation and National Safety Council reports, where Over 50 percent more pedestrians were killed on US roadways in 2021 and in 2022, pedestrian deaths increased about 1\% from 2021 \cite{NSC2024} and \cite{FHWA2022}, where among the frequently involved features are Pedestrian Automatic Emergency Braking, Forward Collision Warning, Traction Control, Auto Reverse System, Lane Departure Warning, and Adaptive Cruise Control.

This upward trend reflects the growing adoption of ADAS-equipped vehicles rather than a direct failure of these systems. The rising involvement of ADAS, however, does raise legitimate issues about effectiveness in real-world operation \cite{RANA2024109237}. Although ADAS is designed to make driving safer, these fatalities reflect possible limitations in system performance, such as difficulty in detecting pedestrians correctly or reacting correctly in difficult situations, and can magnify
potential for driver disengagement and over-reliance on these systems, with their implications for driver behavior, distraction, and road safety.

\subsubsection{Impact of ADAS on Pedestrian Safety}

\textbf{{Internal factors}}

\begin{enumerate}
    \item \textbf{\textit{Travel Speed:}} Fig. \ref{fig:Travel Speed} revealed that fatalities involving ADAS-equipped vehicles tend to occur at slightly lower speeds compared to those without ADAS. Around 40 - 60 mph for both vehicle types, a notable peak is seen; also, ADAS-involved fatalities exhibit a sharper density concentration within this range, suggesting that ADAS-equipped vehicles are more frequently involved in pedestrian fatalities at moderate speeds. Interestingly, on higher speeds (above 80 mph), non-ADAS vehicles have a bit broader density distribution, indicating more pedestrians being killed at high speeds in the lack of ADAS. This may suggest that ADAS is accomplishing some prevention of high-speed pedestrian crashes, but it is not entirely preventing deaths in the mid-range speeds.  
    
\begin{figure}[h!]
\centering
\includegraphics[width=1\textwidth]{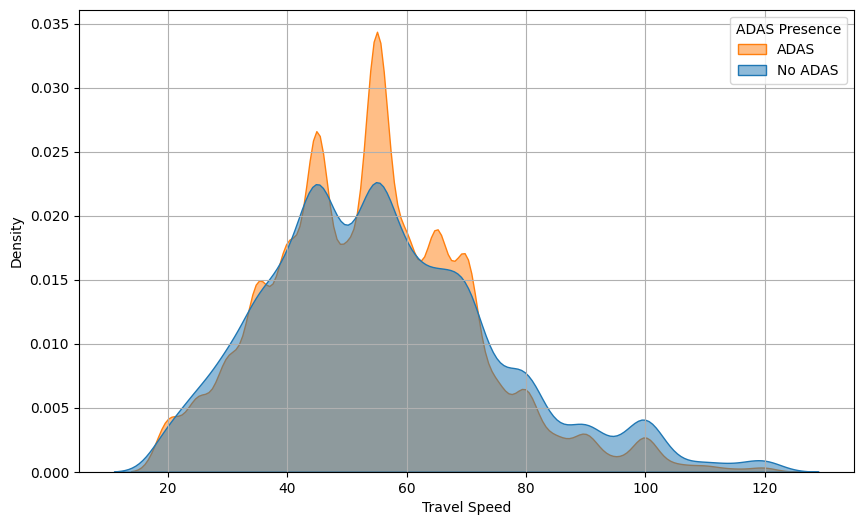}
\caption{Travel speed distribution by ADAS presence.}
  \label{fig:Travel Speed}
\end{figure}

\item \textbf{\textit{Hour of the day:}} Fig. \ref{fig:Hour of the day}; The time-of-day breakdown of pedestrian fatalities differentiates cases with and without ADAS involving pedestrian fatalities. The analysis revealed substantial temporal patterns:

\begin{enumerate}
    \item Late-Night and Early-Morning Hours (12 AM - 6 AM): Non-ADAS pedestrian fatalities have higher concentration during early morning hours, particularly at 2 AM and 5 AM. This could be attributed to low visibility, driver fatigue, or impairment (e.g., drunk driving accidents) \cite{Zhang2016}. Conversely, fatalities with ADAS vehicles are relatively lower during this period.
    \item Day and Afternoon Hours (7 AM - 4 PM): There is a comparable trend for both non-ADAS and ADAS vehicles with fatalities rising uniformly in the morning and peaking later in the afternoon. This suggests that daylight pedestrian fatalities are not significantly impacted by ADAS presence, possibly due to the fact that there are more pedestrian activities in city areas and as a result of more pedestrians with distracting smartphones \cite{Schwebel2022}.
    \item Evening Peak (5 PM - 9 PM): There is a peak pedestrian fatality density at night, with cars with ADAS having a slightly higher density than non-ADAS cars. It occurs during the peak of rush hour traffic, lower natural lighting conditions, and high pedestrian activity. The high fatality rate of ADAS cars during this time period is a problem in the context of ADAS performance at night or in low-light city conditions, when sensor limitations and pedestrian detection issues can arise \cite{10915593}.
    \item Post-9 PM Decline: ADAS as well as non-ADAS vehicles both see decreasing fatality rates after 9 PM, though non-ADAS vehicle collisions have a slightly longer tail.
\end{enumerate}

\begin{figure}
\centering
    \includegraphics[width=1\textwidth]{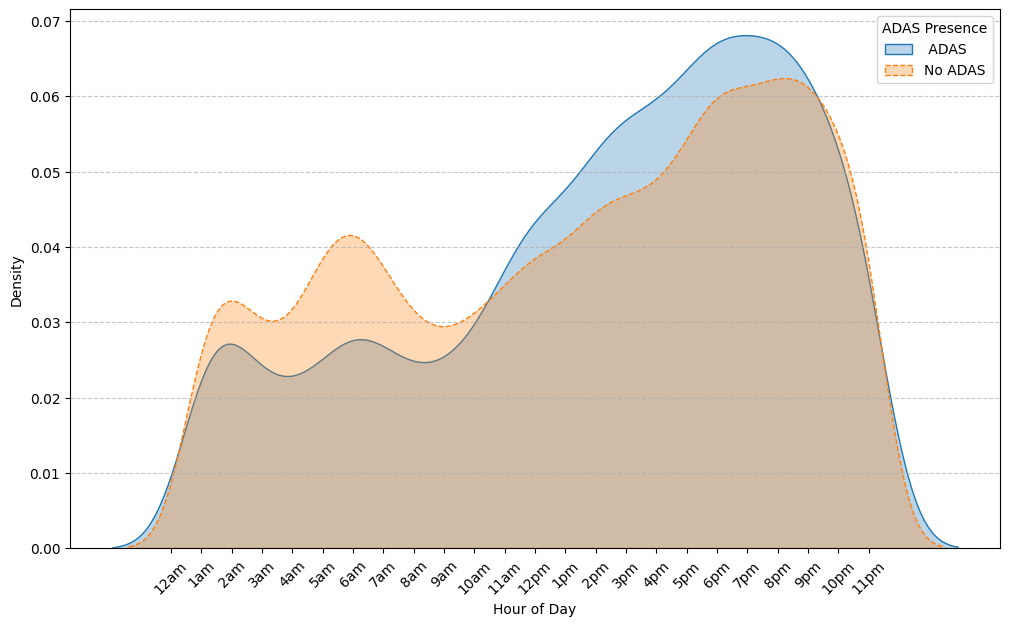}
\caption{Crash distribution by hour and ADAS presence.}
\label{fig:Hour of the day}
\end{figure}

\item \textbf{\textit{Behavioral Contributions to ADAS Performance: }}
Table: \ref{tab:adas_distraction} shows the Analysis of driver-related distractions and driver impairments in crashes involving pedestrians; we observe notable differences between incidents involving ADAS-equipped vehicles and those without ADAS.
\begin{enumerate}
    \item Distracted Driving: Cell phone-related distractions, phone talking and listening, represented 13\% of ADAS vehicle crashes, compared to 87\% of non-ADAS vehicle crashes.External Distractions, like handling a phone, followed the same trend, 11\% of ADAS crashes and 89\% of non-ADAS crashes. In-car object/device-related distractions, such as adjusting controls or inattention due to smoking, were of comparatively larger impact on ADAS vehicles at 17\%, compared to 83\% in the non-ADAS instances, which could indicate restraints in ADAS where the drivers engage in secondary activities within the vehicle. Activities While Driving, eating, drinking, or interacting with passengers, comprised 12\% of ADAS crashes and 88\% of non-ADAS instances. These trends suggest that while ADAS can reduce overall distraction-related crashes, its effectiveness is a function of the type of distraction, particularly when drivers are engaging with in-vehicle devices.
    \item  Physical/Mental State, including alcohol or drug impairment, emotional distress, or physical disabilities, was responsible for 18\% of ADAS-involved crashes and 82\% in non-ADAS cases. Substance Influence, like fatigue or drowsiness, contributed to 14\% of ADAS crashes and 86\% of non-ADAS incidents.
    These findings suggest that while ADAS can mitigate some impairment-related crashes, it is less effective against driver fatigue and physical impairments, likely due to the unpredictable nature of such impairments.
\end{enumerate}

When examining pedestrian behavior contributing to the potential for a crash, we categorized pedestrian activity into two broad categories: Distracted Pedestrians and Impaired Pedestrians and contrasted their history of involvement in crashes with and without Advanced Driver Assistance Systems (ADAS).
\begin{enumerate}

    \item Distracted Pedestrians: The biggest group was Inattention and Distraction, comprising 15\% of the cases where the cars were equipped with ADAS and 85\% of the cases where there was no ADAS.Failure to Comply with Traffic Regulations, like jaywalking and inappropriate crossings, accounted for 8\% of cases with ADAS and 92\% without ADAS. Similarly, Improper Roadway Presence and Visibility Problems—such as people standing or recumbent in the roadway or dark-colored clothing—showed similar patterns, with only 7\% occurring with ADAS, compared to 93\% within non-ADAS environments. 
    This indicates that ADAS-enabled vehicles can provide some level of mitigation against pedestrian errors, yet their capability remains restricted in some distraction-related scenarios. 
    \item Impaired Pedestrian: Substance Influence, e.g., alcohol and drug use, was reported in 13\% of ADAS-equipped vehicle cases and 87\% of non-ADAS cases.Similarly, Physical Impairment, such as the use of a cane or mobility restriction, was the causative factor for 13\% of ADAS incidents and 87\% of non-ADAS. Emotional and Mental State, which comprises states such as fatigue, sickness, or emotional upset, was the causative factor for 12\% of ADAS incidents and 88\% of non-ADAS. These statistics demonstrate that although ADAS would reduce somewhat the risk of pedestrian collisions in such cases, it is not as effective for those entailing cognitive or physical impairment, most likely because there is no warning of such pedestrians' actions.
\end{enumerate}
These numbers suggest that while ADAS might slightly reduce the likelihood of pedestrian crashes in these cases, it is less effective in addressing incidents involving physical or cognitive impairments, likely due to unpredictable pedestrian behavior.

\begin{table}[h]
    \centering
    \renewcommand{\arraystretch}{0.8} 
    \setlength{\tabcolsep}{10pt} 
    \small 
    \begin{tabular}{lcc}
        \toprule
        \textbf{Category} & \textbf{ADAS} & \textbf{No ADAS} \\
        \midrule
        \multicolumn{3}{l}{\textbf{Distracted Driving}} \\
        Cellular Phone Related & 13\% & 87\% \\
        External Distractions & 11\% & 89\% \\
        In-Vehicle Device & 17\% & 83\% \\
        Activities While Driving & 12\% & 88\% \\
        \midrule
        \multicolumn{3}{l}{\textbf{Impaired Driving}} \\
        Physical/Mental State & 18\% & 82\% \\
        Substance Influence & 14\% & 86\% \\
        \midrule
        \multicolumn{3}{l}{\textbf{Distracted Pedestrian}} \\
        Failure to Follow Traffic Rules & 8\% & 92\% \\
        Inattention and Distraction & 15\% & 85\% \\
        Improper Presence in Roadway & 7\% & 93\% \\
        Visibility Issues & 7\% & 93\% \\
        \midrule
        \multicolumn{3}{l}{\textbf{Impaired Pedestrian}} \\
        Substance Influence & 13\% & 87\% \\
        Physical Impairment & 13\% & 87\% \\
        Emotional and Mental State & 12\% & 88\% \\
        \bottomrule
    \end{tabular}
    \caption{Comparison of ADAS vs. No ADAS in various distraction and impairment scenarios for drivers and pedestrians.}
    \label{tab:adas_distraction}
\end{table}

\end{enumerate}

\textbf{{External factors}}
\begin{enumerate}
    \item \textbf{\textit{Lighting Conditions:}} Fig. \ref{fig:Lighting conditions impact ADAS response in pedestrian crashes} shows the comparison of pedestrian fatalities in non-ADAS and ADAS-equipped vehicles and emphasizes the remarkable influence of light conditions on the crash outcome. The pie chart is shown to express that 21.4\%of pedestrian fatalities occurred in ADAS-equipped vehicles, while the most frequent, 78.6\%, occurred in non-ADAS vehicles. This means that although ADAS are increasingly being fitted to more cars, pedestrian deaths are still largely occurring with non-ADAS vehicles. If one looks at light conditions, a greater proportion of ADAS-equipped vehicle fatalities (50.8\%) over non-ADAS vehicles (45.9\%) during daylight suggests that ADAS is more active under better lighting but does not completely eliminate pedestrian crashes. Both vehicles exhibit relatively low numbers of fatalities in the early morning and evening hours, with slightly more accidents by the ADAS-equipped vehicles at dusk (3.0\% vs. 2.2\%) but less at dawn (1.5\% vs. 2.1\%). The most disturbing trend occurs at night when there are high fatalities among pedestrians by both types of vehicles. However, ADAS vehicles have a slightly lower rate of nighttime deaths (44.8\%) compared to non-ADAS vehicles (49.9\%), suggesting that while ADAS may provide some level of nighttime crash prevention, it does not eliminate the risk entirely. 
    
    This analysis points to the potential limitations of ADAS in detecting pedestrians in low-light conditions, a widely established issue for sensor-based technology. Enhancing pedestrian detection through augmented night vision, LiDAR-based identification, and sophisticated emergency braking capabilities can be the solution to reducing pedestrian deaths, especially under low-visibility nighttime conditions\cite{nataprawira2021pedestrian}.

\begin{figure}[H]
\centering
  \includegraphics[width=0.5\textwidth]{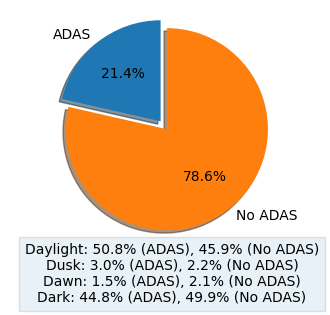}
\caption{Lighting conditions by ADAS status in pedestrian crashes.}
  \label{fig:Lighting conditions impact ADAS response in pedestrian crashes}
\end{figure}%

\begin{figure}[H]
\centering
\includegraphics[width=1\textwidth]{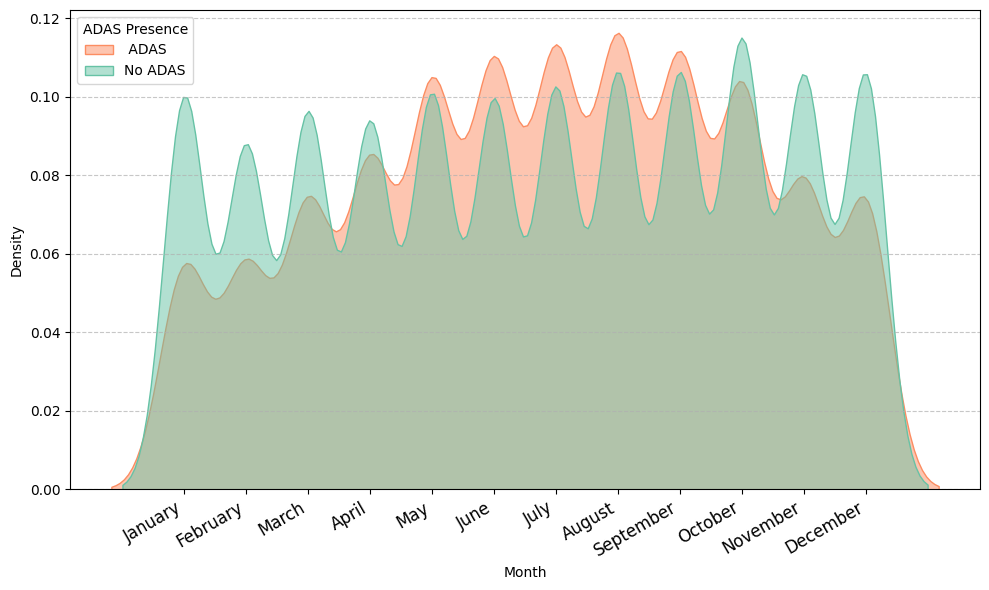}
  \caption{Monthly crash density: ADAS vs. non-ADAS vehicles.}
  \label{fig:Monthly}
\end{figure}

\item \textbf{\textit{Monthly/Seasonal Variations:}} Fig. \ref{fig:Monthly}  illustrates the breakdown of pedestrian fatalities by month from 2018 to 2022, comparing vehicle cases with and without ADAS. The data exhibit a seasonal pattern, with higher pedestrian fatality densities in warmer months (May to October) for both ADAS and non-ADAS vehicles. Interestingly, ADAS vehicles are involved in more pedestrian fatalities during the summer months of June, July, and August, whereas non-ADAS vehicles are involved in more fatalities during winter months, namely February and December. The seasonality of the difference in the number of pedestrian fatalities suggests that increased pedestrian movement during warm months can lead to a larger share of ADAS-equipped vehicle fatalities, possibly due to the presence of more such vehicles on the road. Conversely, non-ADAS fatal crashes with a peak in winter are likely caused by low visibility, slippery roads, or worse driving under poor weather conditions \cite{ahmed2021driver}. The dynamic trends in the density shifts of ADAS versus non-ADAS vehicles over months imply that ADAS could not be equally effective during every season, potentially being plagued by detection challenges during high pedestrian-population density scenarios in summer or low-light conditions in winter weather. These findings indicate that ADAS enhancements, particularly sensor performance in varying environmental conditions, are needed to reduce pedestrian deaths more effectively around the year.

\item \textbf{\textit{Weather Conditions in ADAS performance: }}Fig. \ref{fig:Weather}  highlights the proportion of pedestrian fatalities involving ADAS-equipped and non-ADAS vehicles in different weather conditions. The data indicate that 21.6\% of pedestrian fatalities involved ADAS-equipped vehicles, while 78.4\% involved non-ADAS vehicles, sustaining the pattern that most pedestrian fatalities still occur with non-ADAS vehicles. Analysis of the weather conditions indicates that the majority of the fatalities occurred when it was clear weather, and 80.8\% of the ADAS fatalities and 73.7\% of the non-ADAS fatalities all occurred in clear weather. This suggests that vehicles equipped with ADAS technology cause pedestrian fatalities even in the best of weather, and this is worrisome about their contribution towards prevention of pedestrian collisions in normal driving conditions.
At cloudy conditions, a slightly higher proportion of non-ADAS fatality (15.4\%) compared to ADAS vehicles (13.1\%) was noted, but under rainy conditions, the number of fatalities was greater in non-ADAS vehicles (8.5\%) than ADAS vehicles (4.8\%), showing that although ADAS may minimize impact from crashes in adverse weather, it can never fully eliminate risk. Interestingly, fatalities associated with fog, smog, and smoke were relatively rare but slightly more frequent among non-ADAS vehicles (1.2\%) than among ADAS vehicles (0.9\%). Similarly, snow-related fatalities were relatively more common among non-ADAS vehicles (1.2\%) than among ADAS vehicles (0.4\%), which indicates that ADAS can provide some assistance under adverse visibility and traction conditions.
Overall, these findings suggest that while ADAS may reduce pedestrian fatalities in some adverse weather conditions like rain and snow, it is not so effective in reducing fatalities in favorable weather conditions where most pedestrian accidents still occur \cite{Masello2022}.

\begin{figure}[H]
\centering
  \includegraphics[width=0.5\textwidth]{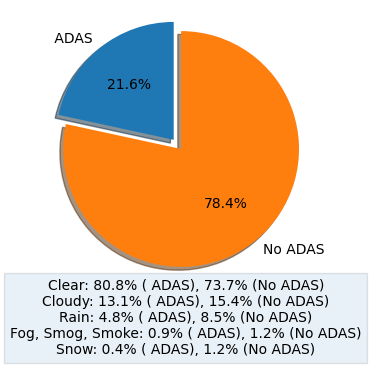}
\caption{Weather conditions affecting ADAS performance.}
  \label{fig:Weather}
\end{figure}%

\begin{figure}[H]
\centering
\includegraphics[width=1\textwidth]{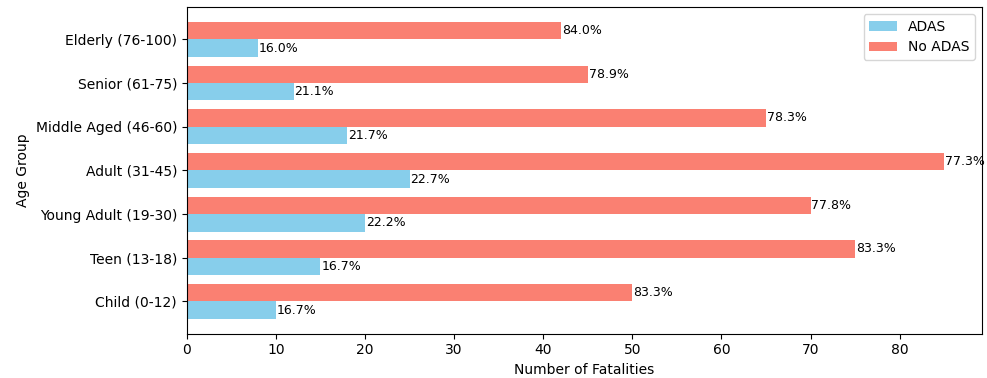}
  \caption{Age group distribution and its relevance to ADAS performance in pedestrian safety.}
  \label{fig:Age-group}
\end{figure}

\item \textbf{\textit{Age-group: }}Fig. \ref{fig:Age-group} presents the distribution of pedestrian fatalities by age group, comparing incidents involving vehicles with and without ADAS. Across all age groups, the majority of fatalities involve non-ADAS vehicles, with percentages ranging from 77.3\% to 84.0\%, reinforcing that pedestrian fatalities predominantly occur with vehicles lacking ADAS. However, a notable share of fatalities (16.0\% to 22.7\%) still involve ADAS-equipped vehicles, highlighting that ADAS does not fully eliminate pedestrian risks.

The oldest age category (61+) has the greatest percentage of fatalities (84.0\%) in collisions with non-ADAS vehicles, which suggests that elderly pedestrians may be more susceptible due to diminished reaction times or mobility problems.
Youth and younger age groups, particularly youth (11-20) and children (0-10), have an 83.3\% fatality rate with non-ADAS vehicles, the same as for young adults (21-30). It is explained by erratic pedestrian behavior, e.g., quick road crossing or distraction.
Of particular note are the middle-aged (31-50) groups that capture the highest percentages of ADAS-related deaths in both cohorts 31-40 and 41-50 at 22.7\% and 21.7\%, respectively. This would indicate that greater proportions of vehicles equipped with ADAS are involved in pedestrian accidents against working-age individuals, possibly due to more pedestrian traffic in urban environments where vehicles equipped with ADAS are more prevalent.

Overall, the evidence shows that even though ADAS could be giving some kind of protection, there are pedestrian fatalities happening at every age category. The higher ADAS-related fatality ratio among middle-aged pedestrians may be an indication of challenges of ADAS detection of pedestrian walkers in dense environments, while the lower ratio for children and elderly citizens suggests that ADAS may be more adept at avoiding collisions with these groups or traditional risk factors (e.g., pedestrian error) remain a dominant effect \cite{SAITO2021221}. These findings emphasize the significance of advanced pedestrian detection and intervention functions in ADAS, particularly in urban environments with higher working-age adult exposure to automobile interaction.

\end{enumerate}

\section{Discussion}

The findings of this study uncover the potential and constraints of Advanced Driver Assistance Systems (ADAS) in reducing pedestrian fatalities. Key results from the efficacy of ADAS in pedestrian safety indicate that certain ADAS components, particularly Pedestrian Automatic Emergency Braking (PAEB) and Forward Collision Warning (FCW), play a major role in preventing pedestrian crashes by identifying oncoming collisions and braking automatically when necessary, a similar phenomenon also registers in \cite{Kidd2024} and \cite{Cicchino2017}. Evidence-based research shows that the automobiles equipped with ADAS are responsible for fewer high-speed pedestrian fatalities than automobiles without ADAS, proving that the systems are effective in controlling crash severity through reduced vehicle speeds and timely alerts to the drivers.
The study further reveals that ADAS has a measurable impact on improving pedestrian safety in clear weather conditions, where optimal visibility allows sensors, cameras, and radar systems to function optimally. In addition, ADAS vehicles demonstrate enhanced crash mitigation performance during the day compared to nighttime, further highlighting the importance of visibility in pedestrian detection \cite{BRUMBELOW2022379}. But the effectiveness of ADAS varies in all instances, and the study identifies certain conditions in which the impact is severely reduced \cite{DEWINKEL2025101336}. Most appreciable limitation exists in low lighting, especially evening, where the sensor-based system of pedestrian detection falters because of minimum contrast and maybe obstructions.At present, owing to advancements, ADAS is still not accurate enough to classify pedestrians from rearview elements of poor light background, resulting in ongoing pedestrian causalities at evening hours \cite{Sucha2021}. In addition, poor weather, such as heavy rain, snow, or fog, worsens the performance of ADAS sensors, decreasing their ability to recognize pedestrians correctly and leading to a system failure threat \cite{ZHANG2023146}.

\subsection{Policy and Industry Implications}
\begin{enumerate}

    \item Integration with Urban Planning: The cities need to incorporate ADAS-friendly infrastructure such as intelligent crosswalks, improved street lights, and networked traffic signals to enhance the accuracy of pedestrian detection, which are important for safer and more efficient roads in urban cities that are rapidly developing with dense traffic. Experiments have proven that smart infrastructure significantly improves the performance of ADAS, particularly in urban cities with dense pedestrian traffic \cite{ELHAMDANI2020102856} and \cite{mitran2020crosswalk}.

    \item Addressing ADAS Limitations in Real-World Scenarios: ADAS operations need to be configured to take into account the different pedestrian behaviors, such as jaywalking and unpredictable movement, in order to enhance pedestrian detection accuracy and reaction time.

     \item Improving Consumer Awareness and Training: Drivers unconsciously overdepend on ADAS, assuming that it can always prevent crashes. Automakers and policymakers need to implement mandatory training programs for drivers to familiarize them with the capabilities and limits of ADAS. Awareness campaigns and interactive in-car reminders will motivate drivers to stay alert and use ADAS optimally \cite{nandavar2023adas}.

      \item Collaboration Between Industry and Policymakers: Automobile manufacturers, innovation technology leaders, and policymakers have to collaborate in order to upgrade ADAS policy from real-world crash statistics and system performance. Wide-ranging tests under different traffic scenarios must be done in order to verify ADAS performance before large-scale rollout. Industry-policymaker cooperation can lead to improved rules, better ADAS performance, and fewer pedestrian fatalities \cite{WANG2023107265}.

\end{enumerate}

\subsection{Limitations}
Inconsistencies in data collection and a lack of standard reports on ADAS activation make it difficult to analyze accurately. ADAS technology experiences difficulties in certain situations, such as low-light conditions and adverse weather conditions. Challenging Pedestrian detection under rain, fog, or snowy conditions, while the intermittent pedestrian movement or urban-level obstacle occurrence can make the system less efficient.

\section{Conclusion and Future Studies}
This study provides a comprehensive, evidence-based examination of the efficacy of Advanced Driver Assistance Systems (ADAS) in reducing pedestrian fatalities, particularly by investigating how different environmental and behavioral factors influence their performance. The findings highlight that even though ADAS technologies, more so Pedestrian Automatic Emergency Braking (PAEB) and Forward Collision Warning (FCW), are extremely contributory to the reduction of crash severity and prevention of some fatalities, their operations are extremely constrained by low-light environments, adverse weather, and random pedestrian movement. Moreover, excessive reliance on the driver and insufficient use of ADAS features are significant problems that render them pointless in terms of safety contribution. The results emphasize that ADAS-equipped vehicles are accountable for a non-trivial percentage of pedestrian fatalities, especially at night and among middle-aged pedestrian populations. These call for improved reliability of sensors, real-time pedestrian detection, and adaptive behavior in varied environments. Manufacturers and policymakers must meet these deficits in aggregate by incorporating ADAS-friendly infrastructure, demanding public education, and calibrating algorithms to take edge-case pedestrian scenarios into account.

Future research should focus on integrating XAI and ADAS for improved transparency and decision-making, especially in more complex urban settings. Beyond the pedestrian-only case. Additionally, elucidates how ADAS functions under multi-agent cases involving cyclists, scooter users, and other vulnerable road users (VRUs).

\section{Credit authorship contribution statement}
\noindent \textbf{Methusela Sulle}: Conceptualization, Writing, and Data analysis. \textbf{Judith Mwakalonge}: Writing, data analysis, methodology, and conceptualization. \textbf{Gurcan Comert}:  Writing, data analysis, methodology, and conceptualization. \textbf{Saidi Siuhi}: Manuscript Preparation. \textbf{Nana Kankam Gyimah}: Manuscript Preparation.

\section{Declaration of competing interest}
The authors declare that no competing financial interests or personal relationships could have influenced the work reported in this paper.

\section{Data availability statement}
This study used a publicly available dataset, and the information is included in this article.

\section{Acknowledgement}
The U.S. Department of Education supported this research through Grant No. P382G320015, administered by the Transportation Program at South Carolina State University (SCSU).


\bibliography{sample}

\end{document}